\documentclass{Interspeech2024}

\usepackage[utf8]{inputenc}
\usepackage{kotex}
\usepackage{booktabs}
\usepackage{caption}
\usepackage{subcaption}
\usepackage{multirow}
\usepackage[hang]{footmisc}
\usepackage[noadjust]{cite}
\usepackage{hyperref}
\interspeechcameraready 

\title{Speech Boosting: Low-Latency  Live Speech Enhancement for TWS Earbuds}

\name[affiliation={1}]{Hanbin}{Bae$^*$}
\name[affiliation={2}]{Pavel}{Andreev$^*$}
\name[affiliation={2}]{Azat}{Saginbaev$^*$}
\name[affiliation={2}]{Nicholas}{Babaev}
\name[affiliation={1}]{Won-Jun}{Lee}
\name[affiliation={1}]{Hosang}{Sung}
\name[affiliation={1}]{Hoon-Young}{Cho}

\address{
  $^{1}$Samsung Research, Seoul, Republic of Korea\\
  $^{2}$Samsung Research \quad $^{*}$equal contribution}
\email{\{bhb0722.bae, p.andreev, a.saginbaev, n.babaev, wall.lee, hosang.sung, h.y.cho\}@samsung.com}

\keywords{low-latency speech enhancement, pruning and quantization, digital signal processor, earbuds, on-device}

\begin{document}

\maketitle

\begin{abstract}
This paper introduces a speech enhancement solution tailored for true wireless stereo (TWS) earbuds on-device usage. 
The solution was specifically designed to support conversations in noisy environments, with active noise cancellation (ANC) activated. 
The primary challenges for speech enhancement models in this context arise from computational complexity that limits on-device usage and latency that must be less than 3 ms to preserve a live conversation. To address these issues, we evaluated several crucial design elements, including the network architecture and domain, design of loss functions, pruning method, and hardware-specific optimization. Consequently, we demonstrated substantial improvements in speech enhancement quality compared with that in baseline models, while simultaneously reducing the computational complexity and algorithmic latency.
\end{abstract}

\section{Introduction}
Recently, true wireless stereo (TWS) earbuds have been successfully popularized along with mobile phones, thereby increasing convenience for many users. In line with this, companies developing earbuds have introduced a variety of functions to maximize user experience. Active noise cancellation (ANC), which blocks almost all sounds around the user, is a core feature of the TWS earbud. This function enhances various experiences in everyday life, such as listening to music, making calls, or focusing on work by removing background noise.

The need for additional technology becomes apparent when a person wearing earbuds and applying ANC wants to have conversations with nearby people. 
Currently, to clearly hear the voice of a nearby person, the user must turn off the ANC function or remove the earbuds altogether. 
If there is a technology that can enhance the voice of a nearby person while reducing ambient sounds through ANC,
this reduces several inconveniences, such as missing a few words, delaying conversation, and increasing the risk of losing earbuds. In this study, we aim to apply a speech enhancement solution focused on advancing the noise suppression capabilities of earbuds, particularly in noisy environments where ANC is in operation. We aim to ensure that suppression does not hinder conversations. This necessitates the development of advanced low-latency speech enhancement models capable of ensuring a balance between noise reduction and critical sound preservation.

To successfully implement an effective speech enhancement model for the aforementioned scenario, two key criteria must be satisfied. First, the algorithmic latency of the speech enhancement module should be maintained at a maximum of 3 ms or less. This is a critical factor in the context of remote communication, where users are noticeably more sensitive to delays. 
This sensitivity arises from the fact that users interact in the same space, making any inconvenience caused by the spectral coloration of the comb-filtering effect~\cite{goehring2018tolerable} from the superposition of the direct and delayed speech more disruptive.
Second, the use of computing resources must be minimized. This is particularly crucial in real-time on-device applications, such as ours, where the efficient use of resources can significantly impact the performance and user experience.

To meet these requirements, we explored several design choices to achieve efficient low-latency speech enhancement.
\begin{enumerate}
    \item We compared the efficiencies of a state-of-the-art frequency-domain network and a time-domain baseline and discovered that the time-domain baseline was more effective when allocated comparable computational resources and algorithmic latency. 
    \item We investigated whether modern structured state space-based models~\cite{goel2022s, gu2021efficiently} could replace our conventional Wave-U-Net + LSTM baseline structure. Despite the encouraging results for long-context modeling tasks, these models were unable to outperform our simple baseline in a low-latency speech enhancement setup. 
    \item We evaluated the efficiency of adversarial losses, a common tool for training contemporary speech enhancement models ~\cite{andreev2022hifi++, shchekotov2022ffc, su2021hifi}, in a low-latency setup and noted its propensity for speech oversuppression. To counter this effect, we suggested two-stage training that combines Phone-Fortified Perceptual Loss (PFPL)~\cite{hsieh2020improving}, adversarial~\cite{andreev2022hifi++}, UTokyo-sarulab Mean Opinion Score (UTMOS)~\cite{saeki2022utmos}, and Perceptual Evaluation Speech Quality (PESQ)~\cite{Audiolabs} losses, that we believe can enhance speech intelligibility and minimize artifacts. 
    \item We assessed the performance of the magnitude pruning method against that of the novel Sparsity Profiles via DYnamic programming search (SPDY) + Optimal Brain Compression (OBC) method~\cite{frantar2022spdy, frantar2022optimal}. We observed that the SPDY + OBC method significantly improved the quality of the pruned models.
\end{enumerate}

Overall, the combination of these techniques delivered low-latency speech enhancement models with a 3 ms algorithmic latency and 0.21 GMAC complexity (or 291 MCPS after being ported on-device), making it suitable for on-device usage while outperforming the baselines with less latency and complexity.

\section{Related Work}
\subsection{Low-latency speech enhancement}
Recently, low-latency speech enhancement has attracted significant research interest. Numerous studies~\cite{tu2021two, zmolikova2021but} have explored time-domain causal neural architectures, such as ConvTasNet~\cite{luo2019conv} and Wave-U-Net + LSTM~\cite{andreev2022iterative}, for this task.

Andreev et al. \cite{andreev2022iterative} suggested a novel training procedure for low-latency models, termed iterative autoregression (IA). This method was used to train a time-domain Wave-U-Net + LSTM model for low-latency speech enhancement, and the benefits of IA were demonstrated in a comparative comparison categorical rating study. In the current study, we examine efficient neural architectures, training losses, and pruning methods. As such, the IA is distinct from the improvements proposed in the IA paper, and we defer the application of the IA to our model for future studies. Another line of works~\cite{wang2022stft,liu2023inplace, cornell2023multi, schroeter2023deep_mf} utilizes time–frequency domain architectures using asymmetric analysis synthesis pairs for windows of short-time Fourier transform or future frame prediction. For example, the iNeuBe-X framework~\cite{cornell2023multi} incorporates the TF-GridNet model~\cite{wang2023tf} for low-latency speech enhancement. The approach based on this architecture won the first place in the 2nd Clarity challenge~\cite{akeroyd20232nd}, a competition for low-latency speech enhancement systems for hearing aids. This method implements low-latency speech enhancement in the time–frequency domain by predicting future short-time fourier transform (STFT) frames, thus reducing the latency imposed by the STFT windows. In the present study, we challenged the TF-GridNet architecture against our time-domain baseline.

\subsection{On-device speech enhancement}

Several studies have been conducted to optimize speech enhancement models for different devices. For TinyLSTMs~\cite{fedorov2020tinylstms}, The authors demonstrated that structural pruning and 8-bit INT quantization could be jointly applied to two LSTM layers. The compressed model achieved a 11.9× reduction in model size and a 2.9× reduction in operations. For the DEMUCS-Mobile~\cite{lee21d_interspeech}, The authors demonstrated that batch normalization pruning utilizing GLU activation could be applied to DEMUCS~\cite{defossez2020real} model. The compressed model achieved a 10x reduction in model size.
However, these models had an algorithmic latency of more than 20 ms that prevented its use in our scenario. Notably, the authors did not study the effect of loss functions on perceptual quality and measured the quality using SI-SDR alone that was poorly correlated with perceptual quality~\cite{andreev2022hifi++}. We also note that with our pruning technique, we managed to reduce the number of operations in our model by 10 times while preserving the high quality, in contrast to 3 times claimed in \cite{fedorov2020tinylstms}. The size of our model achieved less than 1 Mb, in contrast to the 8.9 Mb claimed in \cite{lee21d_interspeech}.

\section{Speech Boosting}
\subsection{Methodology}
\textbf{Data} \ \ \ In all our experiments, we considered additive noise as the distortion to be removed from speech recordings. We employed the VoiceBank-DEMAND dataset~\cite{valentini2017noisy}, a standard benchmark for speech-denoising systems. The training set consisted of 28 speakers with four signal-to-noise ratios (SNR) (15, 10, 5, and 0 dB) and contained 11572 utterances. The test set (824 utterances) consisted of two speakers unseen by the model during training with four SNRs (17.5, 12.5, 7.5, and 2.5 dB). 

\noindent \textbf{Evaluation} \ \ \ We used the state-of-the-art objective speech quality metrics UTMOS~\cite{saeki2022utmos} and WV-MOS~\cite{andreev2022hifi++}. In our internal experiments, we observed that these metrics had the best correlation among well-known objective metrics (for example, SI-SDR, DNSMOS, and PESQ) with human-assigned MOSes for speech enhancement tasks. We also used 5-scale MOS tests for subjective quality evaluation, following the procedure described in \cite{andreev2022hifi++}. 

\begin{figure}[!h]
\vspace{-0.3cm}
  \centering
  \includegraphics[width=0.7\linewidth]{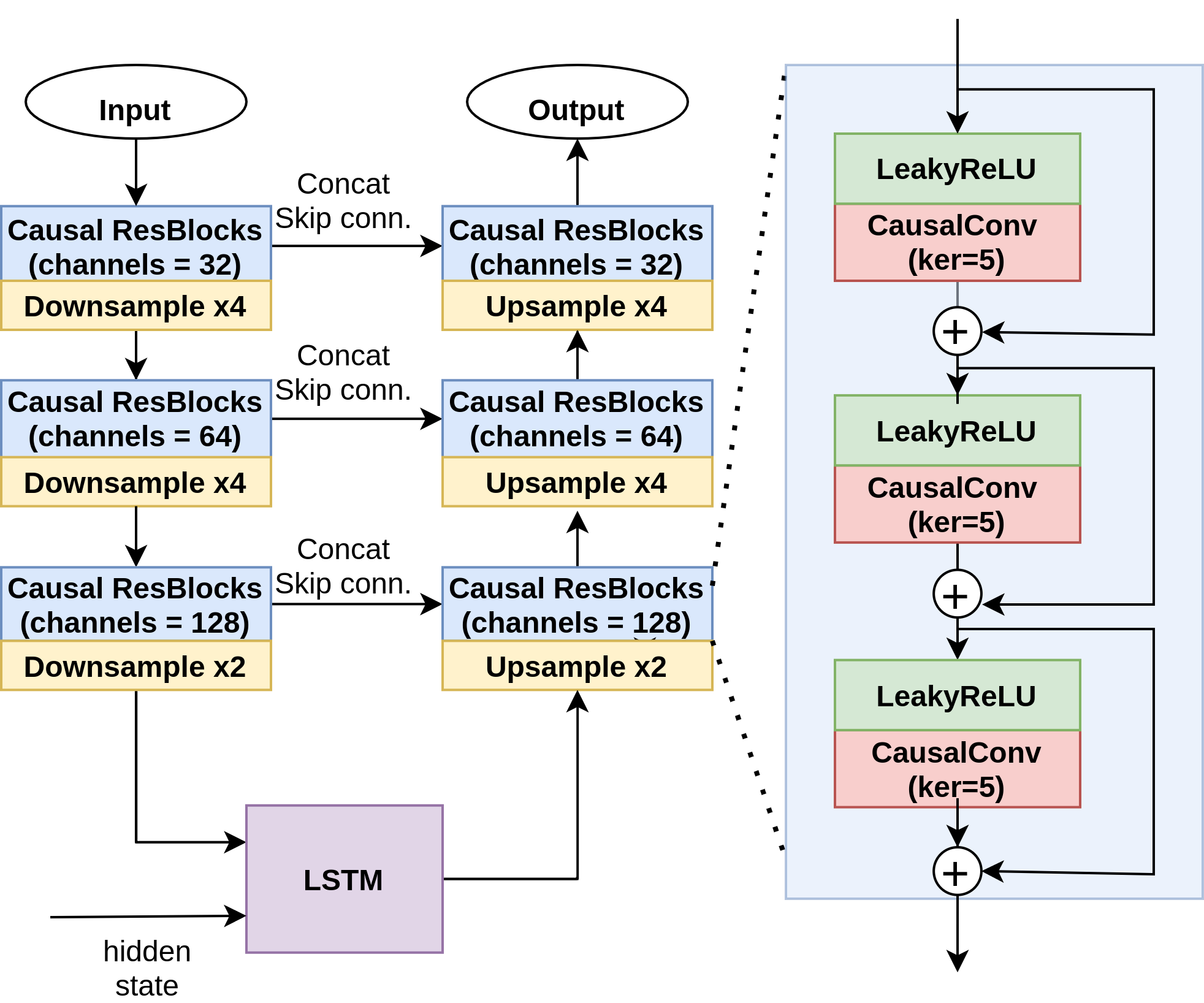}
  \caption{Architecture of the baseline Wave-U-Net + LSTM model.}
  \label{fig:wu}
   \vspace{-0.5cm}
\end{figure}

\noindent \textbf{Baseline} \ \ \   We began with the Wave-U-Net + LSTM baseline introduced in \cite{andreev2022iterative}. We used the adversarial loss function with MSD discriminators because this loss had better perceptual properties than regression-type losses~\cite{su2021hifi, andreev2022hifi++, andreev2022iterative}. We used a three-layer architecture with 4, 4, and 2 strides and 32, 64, and 128 channels, as shown in Figure~\ref{fig:wu}. The model operated at a 16 kHz sample rate. It used a chunk size of 32 timesteps and a look-ahead of 16 timesteps. This look-ahead is achieved by duplicating the input waveform into several channels, each containing a shifted waveform. Consequently, the total algorithmic latency was 48 timesteps, that is equivalent to 3 ms. The computational complexity of this model was approximately 2 GMAC/s. In all the experiments, the batch size was 16, segment size was set to 2 s, and Adam optimizer was used with a learning rate of 0.0002 and betas 0.8 and 0.9.

\subsection{Time domain versus frequency domain}
Many studies have considered time–frequency domain architectures for low-latency speech enhancement. Therefore, our first challenge is to decide whether time or time–frequency domain-based approach is more effective for low-latency speech enhancement. For this purpose, we implemented a state-of-the-art time–frequency domain TF-GridNet architecture~\cite{wang2023tf} for comparison with our Wave-U-Net + LSTM baseline. Because we had to understand whether this time–frequency architecture had benefits over the time-domain baseline, given similar computational constraints, we chose the parameters of the TF-GridNet architecture to match the algorithmic latency to 4 ms and the computational complexity to 2.6 GMAC/s by decreasing the number of channels and TF-GridNet blocks within the network to 4 and 16, respectively. We trained the TF-GridNet model using the same losses as those in the baseline case. The results are summarized in Table~\ref{table:wuvstf}. TF-GridNet appeared to be worse than our baseline. This is likely owing to the high computational complexity of the original TF-GridNet model. The complexity of the original model was approximately 36 GMAC/s, and its performance was significantly degraded when its parameters were reduced, making it impractical for our usage scenario.

\begin{table}[!h]
  \vspace{-0.2cm}
    \centering
    \caption{Time domain model Wave-U-Net + LSTM versus time–frequency domain model TF-GridNet}
      \vspace{-0.1cm}
	\begin{tabular}{l c c c c}
		\toprule
		Model & WV-MOS &  UTMOS & PESQ & \# GMAC  \\
		\midrule
		\midrule
		Ground Truth & 4.50 & 4.32 & 4.26 & - \\
  \midrule
            TF-GridNet & 4.10 & 3.58 & 2.43 &  2.6 \\
             WU+LSTM & 4.30 & 3.81 & 2.55 & 2.0 \\

			\bottomrule
	\end{tabular} \label{table:wuvstf}
	 \vspace{-0.5cm}
\end{table}

\subsection{Model architecture}
In our search for alternative time-domain architectures, we used the recently proposed structured state-space layers (S4)~\cite{gu2021efficiently}. Given the exceptional performance of S4 in sequence modeling tasks, particularly those involving long ranges, we believed that this block could serve as a replacement for the LSTM bottleneck in the Wave-U-Net + LSTM architecture. Consequently, we constructed the Wave-U-Net + S4 architecture as a competitor for our Wave-U-Net + LSTM baseline.
Furthermore, we examined the SaShiMi architecture~\cite{goel2022s} that was built entirely on S4 blocks. SaShiMi leverages the strengths of S4 blocks and is specifically designed for long-range audio modeling tasks. It maintains global coherence by modeling long-range dependencies and incorporates inductive bias through S4 that inherently operates in a continuous-time mode. Given the top-tier performance of SaShiMi in unconditional waveform generation, we hypothesized that it could be a promising candidate to replace our baseline. A comparison between the S4 models and our baseline is presented in Table~\ref{table:sash}. Wave-U-Net + LSTM outperformed the S4 models with similar computational complexities and algorithmic latencies (3 ms).

\begin{table}[!h]
  \vspace{-0.2cm}
    \centering
    \caption{Wave-U-Net + LSTM versus S4 models.}
      \vspace{-0.1cm}
	\begin{tabular}{l c c c c}
		\toprule
		Model & WV-MOS &  UTMOS & PESQ & \# GMAC  \\
		\midrule
		\midrule
            WU+LSTM & 4.30 & 3.81 & 2.55 & 2.0 \\
            	\midrule
            WU+S4&  4.21  & 3.72 & 2.56 & 1.9 \\
            SaShiMi& 4.27 & 3.74 & 2.43 & 2.1 \\
			\bottomrule
	\end{tabular} \label{table:sash}
	 \vspace{-0.5cm}
\end{table}

\subsection{Loss functions}
We observed that the adversarial loss function has two main disadvantages. First, training with this loss is considerably slow because of the training of the discriminators. Second, low-latency models trained with adversarial loss tend to oversuppress the speech content within the recording. This obstacle is expected because the adversarial loss promotes outputs of the model to be within the speech recording distribution rather than preserving the speech content; thus, it may sacrifice some of the speech content to increase the distribution credibility of the generated speech.

Thus, as an alternative, we trained the model using the PFPL~\cite{hsieh2020improving}. The PFPL is a regression-type loss formulated by combining the time-domain L1 loss and the Wasserstein distance between the wav2vec2.0~\cite{baevski2020wav2vec} features of the generated and reference (clean) waveforms.

Usage of the PFPL in the initial training stage offers two significant benefits: (1) This loss impeccably retains speech content during the noise suppression process. This is likely owing to the incorporation of wav2vec2.0 features, known for their proficiency in extracting speech content. (2) Training with the PFPL is considerably faster than with its adversarial counterpart owing to the absence of discriminator training.

However, the use of the PFPL during training sometimes leads to the emergence of background squeak artifacts, a typical phenomenon associated with regression-type losses. To address this, we implemented a second stage of training (fine-tuning) that integrated the adversarial~\cite{andreev2022hifi++}, UTMOS~\cite{saeki2022utmos}, and PESQ~\cite{martin2018deep, kim2019end} losses with 1, 50 and 5 weights, respectively. 

We applied adversarial loss with MSD discriminators as an effective solution for squeak artifacts. This ensured a correlation between the distributions of the clean and generated signals, thereby correcting any distributional discrepancies. Concurrently, UTMOS and PESQ augmented speech intelligibility by incorporating insights gleaned from human preference studies. We utilized the official implementation of the UTMOS score and the PyTorch implementation of the PESQ metric. Both metrics are differentiable with respect to their inputs and can therefore be applied as loss functions (multiplied by negative constants). Owing to the initial stage of PFPL training, we only had to fine-tune the models with second-stage losses for a few epochs, thereby saving time and preserving the speech content captured by the PFPL training.

To verify the efficacy of the proposed training pipeline, we compared it with vanilla adversarial training and vanilla PFPL training. As summarized in the results in Table 3, the proposed two-stage training procedure considerably outperformed the baselines according to human opinion.

\begin{table}[!h]
  \vspace{-0.2cm}
    \centering
    \caption{Comparison of losses. }
      \vspace{-0.1cm}
	\begin{tabular}{l c c c c}
		\toprule
		Loss & WV-MOS &  UTMOS & PESQ & MOS  \\
		\midrule
		\midrule
            Adv. & 4.30  & 3.81 & 2.55 & $3.36\pm0.07$ \\

            PFPL&  4.35 & 3.78 & 2.61 & $3.61\pm0.08$ \\
                    	\midrule
            2-stage&  4.36 & 3.90 & 2.90   & $3.85\pm0.06$ \\
			\bottomrule
	\end{tabular} \label{table:loss}
	 \vspace{-0.5cm}
\end{table}

\subsection{Pruning}
The original Wave-U-Net + LSTM model had a complexity of approximately 2 GMACs, rendering it unsuitable for on-device deployment. We implemented block-structured pruning to optimize the model in terms of performance and storage. 

\begin{enumerate}
    \item For convolutional layers, we applied kernel pruning that enforces sparsity in such a way that if W represents a weight in a convolutional layer, then for certain input channel i and output channel j, \(W[i,j,:] = 0\). We only stored the indices of the non-zero kernels and computed the outputs based on these kernels. 
    \item The LSTM layers were pruned using block sparsity. For each non-zero block, we recorded its coordinates within the fully connected LSTM layers and performed computations only for these non-zero blocks. The blocks measured 16 × 1.
\end{enumerate}

Our pruning pipeline followed an iterative prune + fine-tune strategy. At each pruning iteration, 10\% of the remaining weights were pruned and the model was fine-tuned for 50 epochs. The procedure was continued until the total sparsity of the model $\approx$90\% (complexity-wise). The key is determining the weights required to prune at each iteration. We handled this problem by using the SPDY + OBC pruning strategy. This strategy decomposes the pruning process into layer-wise local pruning (OBC)~\cite{frantar2022optimal} and search for layer sparsity distributions (SPDY)~\cite{frantar2022spdy}.

In the first step of SPDY + OBC, we used the OBC for pruning each layer independently to optimally reconstruct local activations using the mean squared error criterion, given the sparsity constraint. This approach is based on the exact realization of the classical optimal brain surgeon framework applied to local layer pruning. Using the OBC, we obtained a bank of weights for each layer that satisfied different sparsities. 

Subsequently, the SPDY search was employed to determine layer sparsities such that the total model sparsity was suitable for the current computational budget while maximizing the model performance on the calibration data. The algorithm assumed a linear dependency of the model quality on the log-sparsity levels of the layers and used dynamic programming to determine the sparsity levels. The linear dependency parameters were optimized using differential evolution and random search (shrinking neighborhood local search) algorithms for global optimization. 

We compared the proposed pruning pipeline with common baseline magnitude pruning and observed that SPDY + OBC pruning drastically improved the quality of the pruned models under similar complexity constraints.

\begin{table}[!h]
  \vspace{-0.2cm}
    \centering
    \caption{Comparison of pruning methods. }
      \vspace{-0.1cm}
	\label{tab:phase}
	\begin{tabular}{l c c c c}
		\toprule
		Method & WV-MOS &  UTMOS & PESQ &GMAC  \\
		\midrule
		\midrule
            Base model&  4.36 & 3.90 & 2.90 &  2.0 \\
            	\midrule
            Mag. pruning  & 4.09 & 3.63 & 2.62 & 0.29 \\
            SPDY+OBC&  4.27 & 3.90 & 3.01 & 0.21 \\
			\bottomrule
	\end{tabular} \label{table:vocod}
	 \vspace{-0.5cm}
\end{table}

\subsection{HiFi4 DSP simulation}

We implemented the 0.21 GMAC pruned model in the native C code. Running this code on on a system with Cadence Tensilica HiFi4
DSP core~\cite{hifi4dsp} provided 2031 million clocks per second (MCPS) for this model. This is the total number of clocks, including the instructions to load and store each variable required for the calculations through the data memory interface. The clock frequencies supported by the micro control units are typically approximately 300–600 MHz. Because the processing time is longer than the algorithmic latency, delays are inevitable. Therefore, single instruction multiple data (SIMD) operations, such as the 16-bit four-way SIMD operation of HiFi4 DSP for fixed-point numbers, have to be used to reduce the total MCPS. We converted the inputs and parameters of each layer into fixed-point numbers using Q format. In this case, the input values were converted to Q12 as a 32-bit integer variable. The weights and biases of the convolutional layers were converted to Q13 and Q25 as 16-bit short and 32-bit integer variables, respectively. The weights and biases of the LSTM layers were converted to Q13 as 16-bit short variables. Subsequently, we replaced the calculation expressions of the convolutional and LSTM layers with SIMD operations of the HiFi4 DSP. The final optimized model had 291 MCPS and around 800 kB size.

\section{Results}
\subsection{Comparison with existing approaches}

We compared the resulting speech-boosting models with Wave-U-Net + LSTM trained using IA~\cite{andreev2022iterative} and a non-causal DEMUCS denoiser~\cite{defossez2020real}. For the IA baseline, we used Wave-U-Net + LSTM with the K, N, and C parameters set to 7, 4, and [16, 24, 32, 48, 64, 96, 128], respectively. This configuration corresponds to 8 ms of algorithmic latency and 2.0 GMAC complexity. The DEMUCS denoiser was used in a non-causal configuration with the H parameter set to 64. Both these baseline models had considerably higher computational complexity and algorithmic latency than our pruned model, while delivering comparable perceptual quality according to the MOS score.

\newcolumntype{H}{>{\setbox0=\hbox\bgroup}c<{\egroup}@{}}
\begin{table}[!h]
  \vspace{-0.2cm}
    \centering
    \caption{Comparison with DEMUCS and IA.}
      \vspace{-0.1cm}
	\begin{tabular}{l c c H c}
		\toprule
		Model & GMAC &  Alg. latency & UTMOS & MOS  \\
		\midrule
		\midrule
            Input & - & - & - & $3.33\pm0.07$\\

  		\midrule
            Ours & 2.0 & 3 ms & 3.90 & $3.85\pm0.06$\\
            Ours (pruned) & 0.21 & 3 ms & 3.90 & $3.71\pm0.05$ \\
            	\midrule
            DEMUCS & 38.1 & non-caus. & 3.93 & $3.75\pm0.06$ \\
            IA WU+LSTM& 2.0  & 8 ms & 3.74 & $3.77\pm0.05$ \\
			\bottomrule
	\end{tabular} \label{table:vocod}
\end{table}

\begin{figure}[!h]
  \centering
  \includegraphics[width=0.8\linewidth]{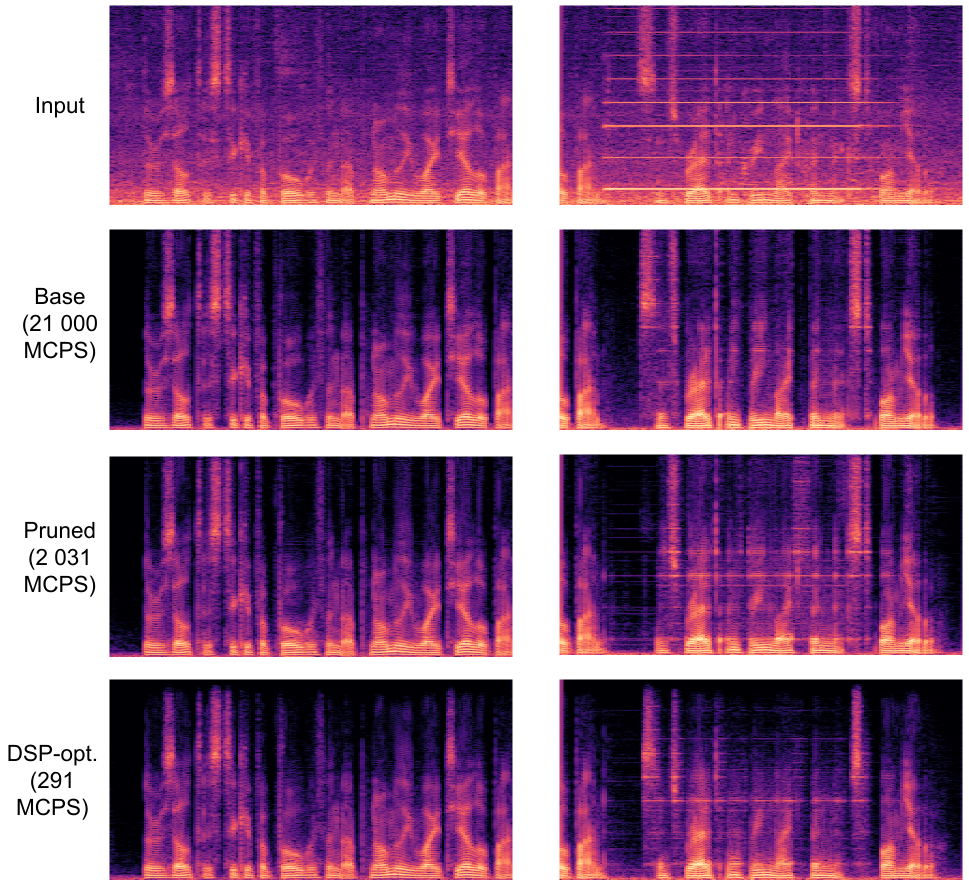}
  \caption{Examples of speech denoising performance.}
  \label{fig:spec}
   \vspace{-0.5cm}
\end{figure}

\subsection{Limitations and future work}
Figure~\ref{fig:spec} shows the enhanced speech samples of all the process models, where the input sample is a male voice mixed with subway noise at an SNR of 2.5 dB. 
Samples mixed with babble noise is shown in the left segment, and samples mixed with the additional harmonic noise of the alarm sound is shown in the right segment.
As observed, the babble noise was well removed, whereas the harmonic noise remained in small amounts in the utterance segments of the enhanced speech. This observation suggests that Wave-U-Net + LSTM struggles to filter out the harmonic signals in the noisy speech adequately. 

Owing to the inherent complexities of speech signals and noise characteristics, accurately estimating and removing noise while preserving the speech components can be a delicate balance. 
In addition, variations in harmonic gains and fluctuations in the denoising process can contribute to the generation of harmonic noise artifacts, making it challenging to achieve a clean and natural sounding output~\cite{cho2012icassp}. 
To alleviate this issue, Wave-U-Net + LSTM should be improved to capture harmonic relationships in noisy speech. 
One promising avenue for future research in this area includes hybrid architectures that operate simultaneously in time and frequency domains~\cite{defossez2022hybrid, rouard2023hybrid, andreev2022hifi++}.

\section{Conclusion}
This work advances low-latency, on-device speech enhancement by reevaluating several critical design choices. We examine different model architectures, training losses, and pruning techniques, selecting the optimal scenario for efficient low-latency speech enhancement. 
The resulting model achieves a remarkable balance between performance and resource utilization. It is suitable for on-device usage, exhibits low algorithmic delay, and delivers a quality comparable to models with significantly higher algorithmic latency.
The experimental results will pave the way for future advancements in speech enhancement technology for TWS earbuds that will be co-operated with various audio processing modules such as ANC and Beamforming.

\bibliographystyle{IEEEtran}
\bibliography{mybib}

\end{document}